\documentclass[aps,prb,twocolumn,superscriptaddress]{revtex4}
\usepackage{graphicx}

\begin{document}

\title{Doping evolution of antiferromagnetism and transport properties in the non-superconducting BaFe$_{2-2x}$Ni$_{x}$Cr$_{x}$As$_{2}$}

\author{Rui Zhang}
\thanks{These authors made equal contributions to this paper}
\affiliation{Beijing National Laboratory for Condensed Matter
Physics, Institute of Physics, Chinese Academy of Sciences, Beijing
100190, China}
\author{Dongliang Gong}
\thanks{These authors made equal contributions to this paper}
\affiliation{Beijing National Laboratory for Condensed Matter
Physics, Institute of Physics, Chinese Academy of Sciences, Beijing
100190, China}
\author{Xingye Lu}
\affiliation{Beijing National Laboratory for Condensed Matter
Physics, Institute of Physics, Chinese Academy of Sciences, Beijing
100190, China}
\author{Shiliang Li}
\affiliation{Beijing National Laboratory for Condensed Matter
Physics, Institute of Physics, Chinese Academy of Sciences, Beijing
100190, China}
\affiliation{Collaborative Innovation Center of Quantum Matter, Beijing, China}
\author{Mark Laver}
\affiliation{Laboratory for Neutron Scattering and Imaging, Paul Scherrer Institute, CH-5232 Villigen, Switzerland}
\author{Christof Niedermayer}
\affiliation{Laboratory for Neutron Scattering and Imaging, Paul Scherrer Institute, CH-5232 Villigen, Switzerland}
\author{Sergey Danilkin}
\affiliation{Bragg Institute, Australian Nuclear Science and
Technology Organization, New Illawarra Road, Lucas Heights NSW-2234
Australia}
\author{Guochu Deng}
\affiliation{Bragg Institute, Australian Nuclear Science and
Technology Organization, New Illawarra Road, Lucas Heights NSW-2234
Australia}
\author{Pengcheng Dai}

\affiliation{Department of Physics and Astronomy, Rice University, Houston, Texas 77005, USA}
\affiliation{Beijing National Laboratory for Condensed Matter Physics, Institute of Physics,
Chinese Academy of Sciences, Beijing 100190, China}

\author{Huiqian Luo}
\email{hqluo@iphy.ac.cn}
\affiliation{Beijing National Laboratory for Condensed Matter
Physics, Institute of Physics, Chinese Academy of Sciences, Beijing
100190, China}

\begin{abstract}
We report elastic neutron scattering and transport measurements
on the Ni and Cr equivalently doped iron pnictide
BaFe$_{2-2x}$Ni$_{x}$Cr$_{x}$As$_{2}$. Compared with the electron-doped
BaFe$_{2-x}$Ni$_{x}$As$_{2}$,
the long-range antiferromagnetic (AF) order
in BaFe$_{2-2x}$Ni$_{x}$Cr$_{x}$As$_{2}$
is gradually suppressed with vanishing ordered moment and N\'{e}el temperature
near $x= 0.20$ without the appearance of superconductivity.
A detailed analysis on the transport properties of BaFe$_{2-x}$Ni$_{x}$As and BaFe$_{2-2x}$Ni$_{x}$Cr$_{x}$As$_{2}$
suggests that the non-Fermi-liquid behavior associated with the linear resistivity as a function of temperature
may not correspond to the disappearance of the static AF order.
From the temperature dependence of the resistivity in overdoped compounds without static AF order, we find that
 the transport properties are actually affected by Cr impurity scattering, which may induce a
metal-to-insulator crossover in highly doped BaFe$_{1.7-y}$Ni$_{0.3}$Cr$_{y}$As$_{2}$.

\end{abstract}


 \maketitle

\section{Introduction}

A determination of the complex phase diagram in high-transition (high-$T_c$) temperature superconductors is the first step
to understand the mechanism of high-$T_c$ superconductivity \cite{PALee,scalapino,Keimer,Coleman}.
In iron pnictides, the parent compounds exhibit a tetragonal-to-orthorhombic lattice distortion below $T_s$ and
antiferromagnetic (AF) order below $T_N$. While superconductivity (SC) emerges upon sufficient electron/hole or isoelectronic doping to the parent compounds
such as BaFe$_2$As$_2$ \cite{kamihara,cruz,stewart,pdai},
the gradual suppression of
the orthorhombic lattice distortion and AF order near optimal superconductivity suggests the presence of a
quantum critical point (QCP) hidden beneath the superconducting dome, which may be responsible for the anomalous normal-state properties and the high-$T_c$ superconductivity \cite{Shibauchi,Zaanen}.
However, systematic neutron scattering (NS) and X-ray diffraction(XRD) experiments on the electron doped
BaFe$_{2-x}$Ni$_x$As$_2$ \cite{Rotter,assefat,ljli,qhuang,nni} reveal that while $T_s$ and $T_N$ initially decrease with increasing electron-doping, their values saturate near optimal superconductivity associated with an incommensurate AF order, resulting in an avoided magnetic QCP \cite{hqluo2012,xylu2013}. In particular, the short-ranged incommensurate AF ordered phase near optimal superconductivity is consistent with a cluster spin glass phase in the matrix of
the superconducting phase \cite{xylu2014}. Similar conclusions are also reached
from muon spin relaxation ($\mu$SR) \cite{Bernhard} and Nuclear magnetic resonance (NMR) measurements \cite{dioguardi} on
BaFe$_{2-x}$Co$_x$As$_2$ family of materials \cite{pratt,kim}, while a separate NMR measurement
on BaFe$_{2-x}$Ni$_x$As$_2$ suggests the presence of a magnetic QCP at $x$= 0.10 and a structural QCP at $x$= 0.14,
both associated with the non-Fermi-liquid behavior \cite{rzhou}.
For iso-valently doped BaFe$_{2}$(As$_{1-x}$P$_x$)$_2$, the quantum critical behavior has been reported near optimal superconductivity
around $x=0.3$ from transport and superfluid density measurements \cite{Hashimoto,sjiang,Kasahara,Analytis}.  However, recent
neutron diffraction measurements show coincident $T_s$ and $T_N$ in the underdoped regime, and the linear extrapolation of
the orthorhombic lattice distortion suggests that structural QCP cannot exceed $x$= 0.28 doping\cite{Allred}.  Similarly, it is still unclear if there is a QCP in the
hole-doped Ba$_{1-x}$K$_x$Fe$_2$As$_2$ \cite{Avci}.  To summarize, there are currently no direct evidence demonstrating a gradual suppression of the
$T_N$ and $T_s$ to zero temperature with increasing doping inside the superconducting dome.  Moreover,
it is unclear how magnetism coexists and competes with superconductivity near optimal doping.  This is because the occurrence of superconductivity tends
to suppress the static AF order and orthorhombic lattice distortion \cite{hqluo2012,xylu2013,Bernhard,pratt,kim}, thus complicating the ability to determine the intrinsic nature of the
AF order without the influence of superconductivity.

\begin{figure}[t]
\includegraphics[width=0.4\textwidth]{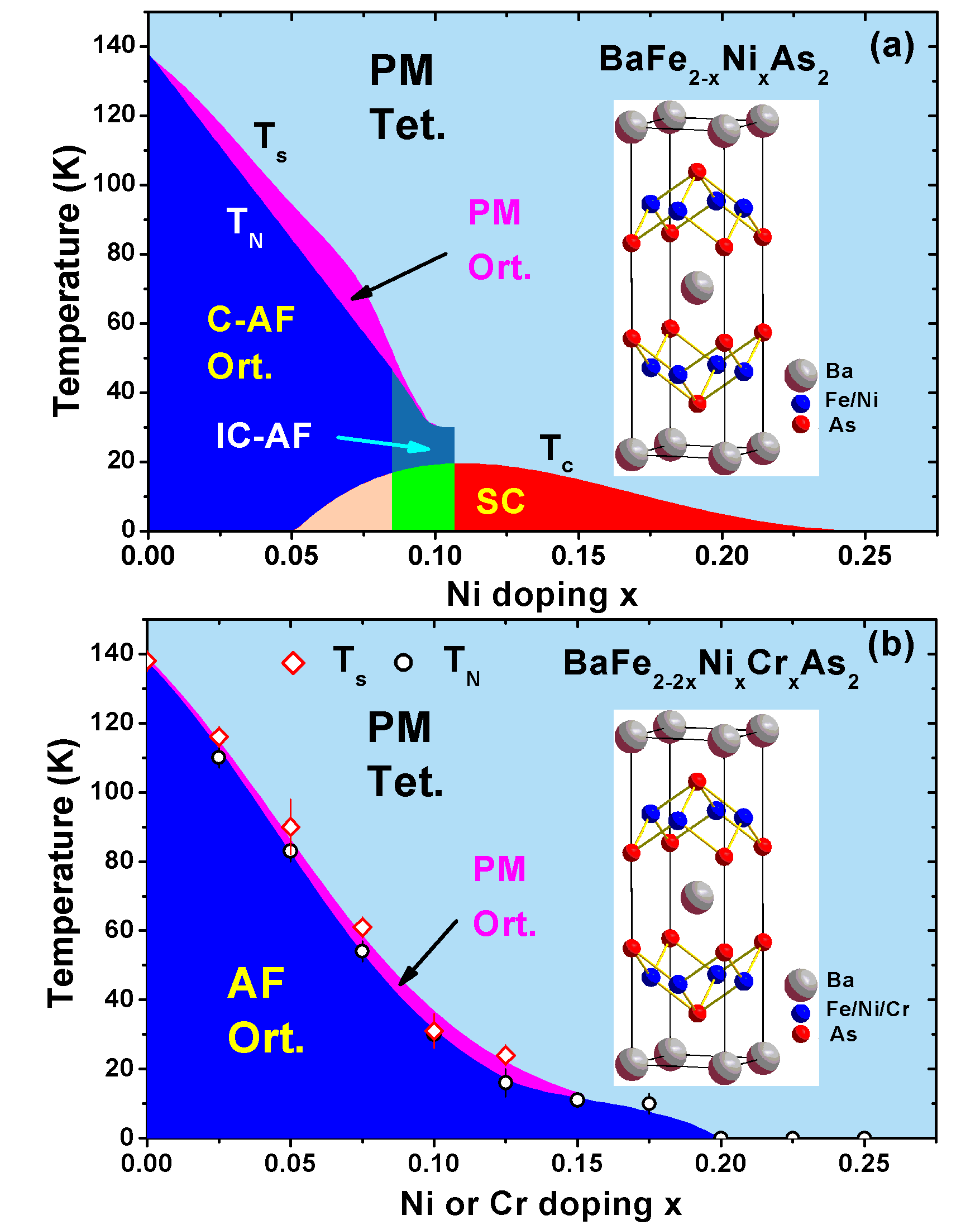}
\caption{(Color online) The electronic phase diagram of (a) BaFe$_{2-x}$Ni$_{x}$As$_{2}$ and
(b) BaFe$_{2-2x}$Ni$_{x}$Cr$_{x}$As$_{2}$. Upon cooling in both systems, a structure
transition from tetragonal (Tet.) to orthorhombic (Ort.) lattice and
a magnetic transition from paramagnetism (PM) to antiferromagnetism
(AF) occur at $T_s$ and $T_N$, respectively.  By approaching the optimal superconductivity around $x= 0.10$ in BaFe$_{2-x}$Ni$_{x}$As$_{2}$, The long-ranged commensurate antiferromegnetism (C-AF) degenerates into short-ranged incommensurate antiferromegnetism (IC-AF). Insert shows the
crystal structure. }
\end{figure}

One way to determine if there is indeed a magnetic QCP upon
suppression of the static AF order is to carry out experiments on
materials without superconductivity.  For example, transport and
neutron powder diffraction experiments on CeFeAs$_{1-x}$P$_x$O
reveals the presence of a magnetic QCP near $x=0.4$ without
superconductivity that is controlled by the pnictogen height (the
average Fe-As/P distance) away from the iron plane
\cite{Cruz2010,YKLuo2010}. Since impurity doping onto the FeAs plane
for iron pnictides will suppress superconductivity
\cite{pcheng2010,yfguo,jli2011,jli2012,jli2014,pcheng2013,ykli,tinabe,qdeng},
it will be interesting to find a system where one can gradually
reduce the AF order with increasing doping but without inducing
superconductivity. We have found that Cr impurity doping into
BaFe$_{2-x}$Ni$_x$As$_2$ is very efficient in suppressing
superconductivity without much affecting $T_s$ and $T_N$ of the
materials \cite{rzhang}, it would be interesting to carry
out neutron scattering NS experiments on these materials to determine
the evolution of the structural and magnetic phase transitions in
Cr-doped BaFe$_{2-x}$Ni$_x$As$_2$ system without the effect of
superconductivity.

In this article, we report the NS and electric transport studies on the Ni and Cr equivalently doped iron pnictide BaFe$_{2-2x}$Ni$_{x}$Cr$_{x}$As$_{2}$.  Compared with the pure Ni-doped BaFe$_{2-x}$Ni$_{x}$As$_{2}$ system [Fig. 1 (a)] \cite{hqluo2012,xylu2013,xylu2014},
we find no evidence for superconductivity while the static AF order is gradually suppressed with increasing equal amounts of Ni- and Cr- doping [Fig. 1 (b)].  For $x\leq$ 0.10,
$T_s$ and $T_N$ are almost the same as Cr free samples.  Upon further increasing $x$,
$T_N$ decreases continuously until vanishes (within our sensitivity) above $T=1.5$ K for
 $x \geq 0.20$.  From analysis of the resistivity data at temperature above $T_N$, we find
no dramatic anomaly for transport behavior around the AF zone boundary $x$= 0.20.
Similar resistivity analysis in BaFe$_{2-x}$Ni$_{x}$As$_{2}$ suggests possible
non-Fermi-liquid behavior in the overdoped regime around $x$= 0.15. By fixing Ni composition at $x$= 0.30, increasing Cr doping may significantly affect the transport properties by introducing Kondo or weak localization effects due to impurity scattering, resulting in a possible metal-to-insulator crossover in highly doped BaFe$_{1.7-y}$Ni$_{0.3}$Cr$_{y}$As$_{2}$ ($y\geq$ 0.30). Our results suggest that the non-Fermi-liquid behavior determined from transport measurement is not intimately associated with the disappearance of magnetic order or lattice orthorhombicity,
and the impurity scattering is important to the transport properties of iron pnictides.

\begin{figure}[t]
\includegraphics[width=0.4\textwidth]{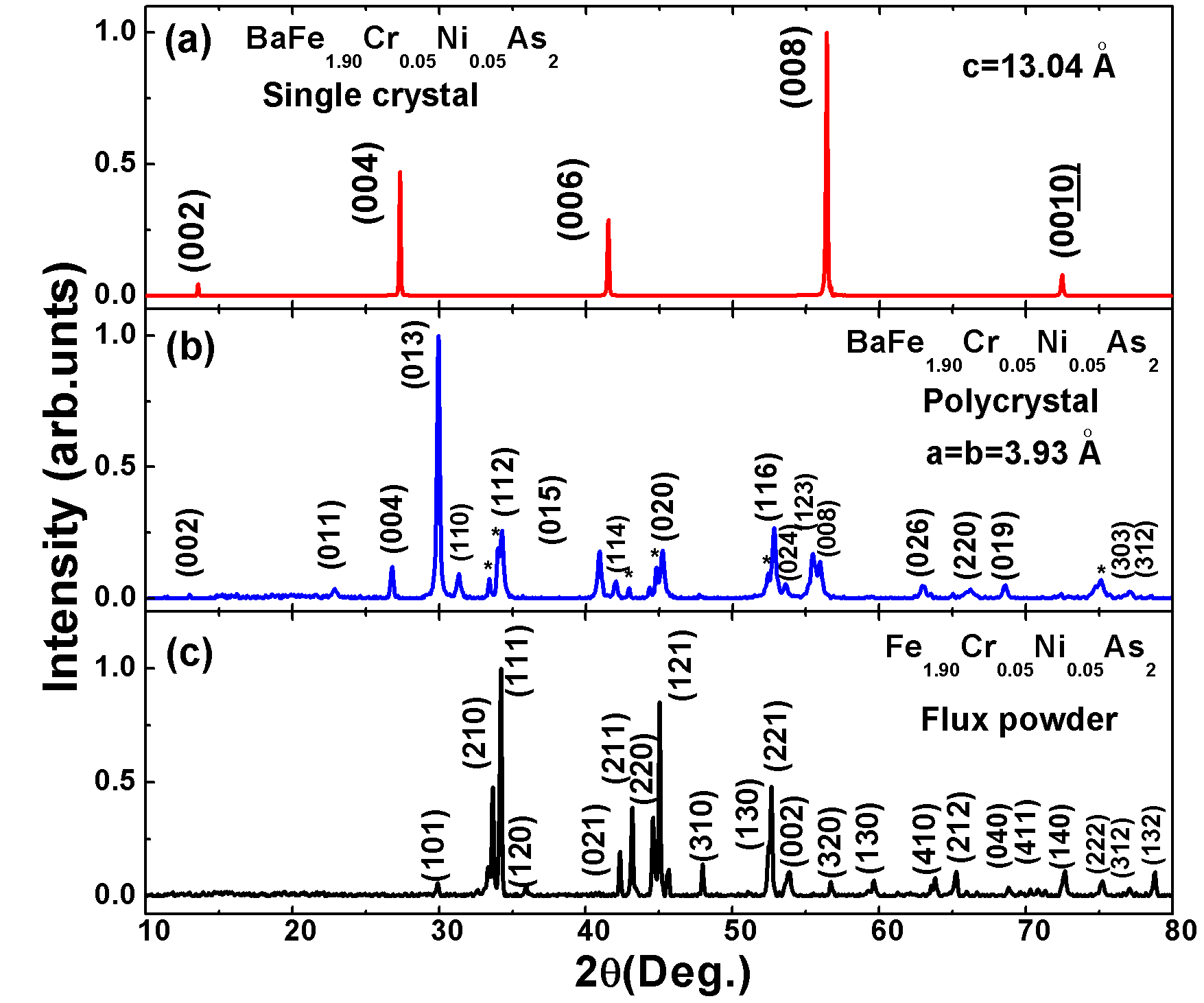}
\caption{(Color online) Typical patterns of X-ray diffraction on the
(a) as-grown single crystal, (b) polycrystal crushed from single
crystal of BaFe$_{1.9}$Ni$_{0.05}$Cr$_{0.05}$As$_{2}$, and (c) flux
powder of Fe$_{1.9}$Ni$_{0.05}$Cr$_{0.05}$As$_{2}$. The stars mark
the peaks from residual flux in the crushed crystal. For clarity,
the intensity is normalized to [0, 1].
 }
 \end{figure}

\section{Experiment}
Single crystals of BaFe$_{2-2x}$Ni$_{x}$Cr$_{x}$As$_{2}$ were grown by self-flux method similar to BaFe$_{2-x}$Ni$_{x}$As$_{2}$ \cite{ychen,rzhang}.
The in-plane resistivity ($\rho_{ab}$) was measured by the standard four-probe method in a \emph{Quantum Design} Physical Property Measurement System. All measured crystals were cut into rectangluar shapes $4\times 0.5\times0.1 $ mm$^3$ with the $c$-axis being the smallest dimension. Four Ohmic contacts with low resistance (less than 1 $ \Omega$) on $ab$-plane were made by silver epoxy.  In order to lower noises, a large current $I=5$ mA and slow sweeping rate of temperature (2 K/min) were applied. Measurements on each doping were repeated on 3 $\sim$ 5 pieces of crystals to confirm the temperature dependence of $\rho_{ab}(T)$ and reduce uncertainty in estimation of the absolute resistivity due to geometric factors. By directly comparing the temperature dependence of resistivity at different doping concentrations, we normalized the resistivity $\rho_{ab} (T)$ by the data at room temperature ($T=300$ K).

We carried out elastic neutron scattering experiments using the Rita-2 triple-axis spectrometers at Swiss Spallation Neutron Source, Paul Scherrer Institute, Switzerland.
To eliminate the scattering from higher-order neutrons with wavelength $\lambda /n (n \geq 2)$ , a pyrolytic graphite (PG) filter before the sample and a
cold Be filter after the sample were used. The fixed final energy was $E_f=4.6$ meV with a wave length $\lambda_f=4.2 $ \AA.
The collimation for supermirror guide after the monochromator was $80^\prime$, the radial collimator after the Be filter was about $150^\prime$, the effective collimation of the neutron-absorbing guide after
the analyzer was $40^\prime$ for preventing cross talk between the different detector channels. We also took data on the $x=0.175, 0.20, 0.25$ compounds using the TAIPAN thermal neutron triple-axis spectrometer at the Bragg Institute, Australian Nuclear Science and Technology Organization (ANSTO). The PG filters were placed before and after the sample. To get better resolution, the PG monochromator and analyzer were set to the flat mode without collimation. The fixed final energy was selected as $E_f=14.87$ meV ($\lambda_f=2.3 $ \AA) or $9.03$ meV ($\lambda_f=3 $ \AA).  We define the wave vector $\bf Q$ at ($q_x$, $q_y$, $q_z$) as $(H,K,L) = (q_xa/2\pi, q_yb/2\pi, q_zc/2\pi)$ reciprocal lattice units (r.l.u.) using the tetragonal lattice parameters $a \approx b \approx 3.94$ \AA, and $c \approx 12.90$ \AA.  For each experiment, a single crystal with mass of nearly 1 gram was aligned to the $[H, H, 0] \times [0, 0,L]$ scattering plane. The thickness of our sample is about 0.5 mm, and the neutron absorption is negligible due to small neutron absorption cross sections for all the elements.

\begin{figure}[t]
\includegraphics[width=0.45\textwidth]{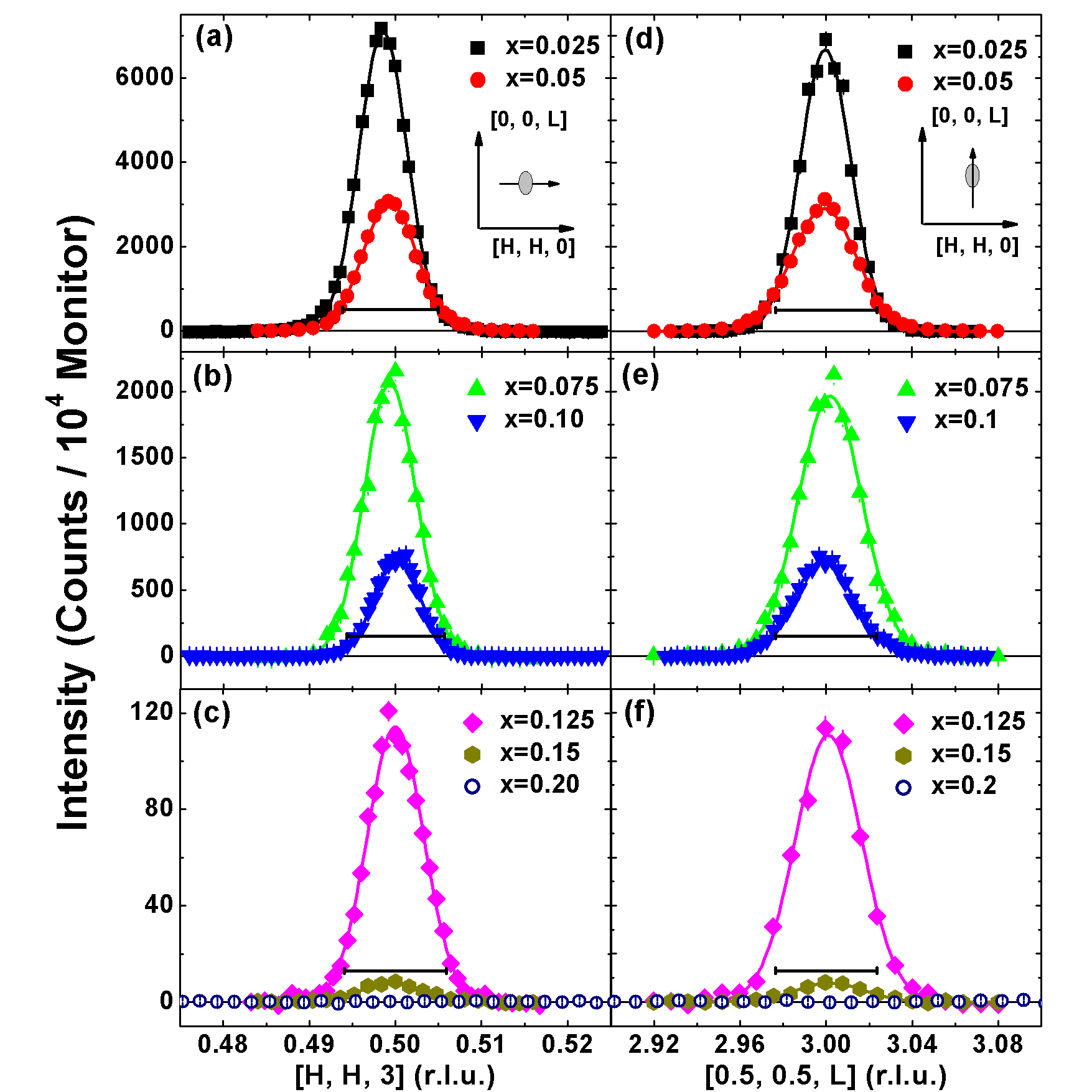}
\caption{(Color online) ${\bf Q}$ scans for the antiferromagnetic
peak at (0.5, 0.5, 3) along (a) - (c) $[H, H, 3]$ and (d) - (f)
$[0.5, 0.5, L]$ at 2 K. The black horizontal bars are the
instrumental resolution determined by using $\lambda/2$ scattering
from the
 $(1, 1, 6)$ nuclear Bragg peak above $T_N$ without filter. All data are subtracted by
the background above $T_N$. The solid lines are gaussian fitting results.}
\end{figure}

\begin{figure*}[t]
\includegraphics[width=0.6\textwidth]{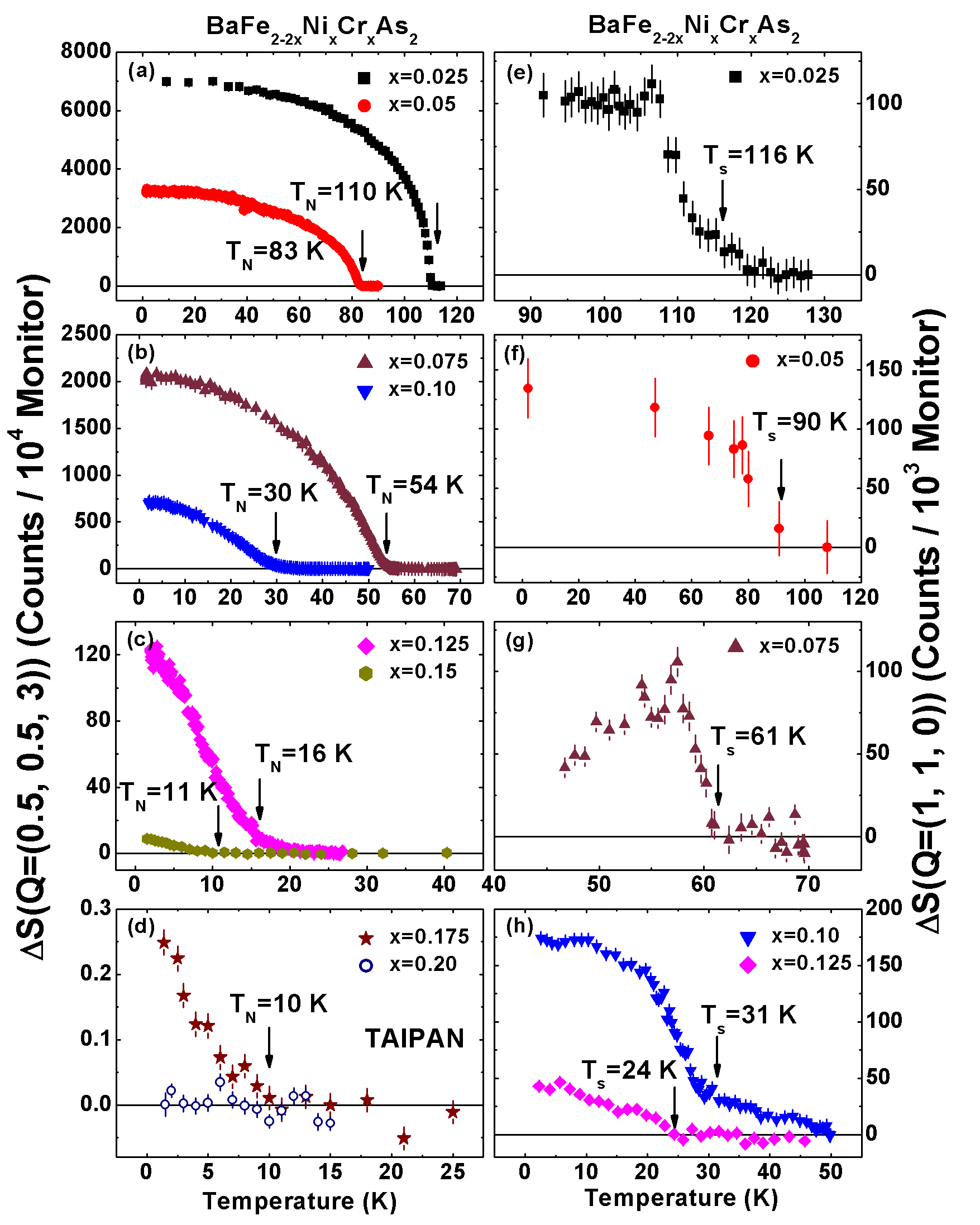}
\caption{(Color online) (a) - (d) Order parameter of
antiferromagnetism at $Q_{\mathrm{AF}}=(0.5, 0.5, 3)$. All data are
subtracted by the background above $T_N$ and normalized to $10^4$
monitor counts.
 (e) - (h) Order parameter of structure
transition from neutron extinction effect at $Q=(1, 1, 0)$.  All
data are subtracted by the background above $T_s$ and normalized to
$10^3$ monitor counts.}
\end{figure*}

\section{Results and Discussion}

\subsection{Sample characterization}

We first describe the sample characterization of BaFe$_{2-2x}$Ni$_{x}$Cr$_{x}$As$_{2}$ single crystals.  Figure 2 (a) shows the XRD patterns on one typical crystal BaFe$_{1.90}$Ni$_{0.05}$Cr$_{0.05}$As$_{2}$. The sharp even peaks $(0 0 l)$ ($l=2, 4, 6, ...$) along the $c$-axis with narrow width about 0.1$^\circ$ indicate high crystalline quality of our sample. The calculated $c-$axis parameter is about 13.04 \AA\ for this sample, closed to the parent compound BaFe$_{2}$As$_{2}$. Fig. 2 (c) shows the XRD results for Fe$_{1.90}$Ni$_{0.05}$Cr$_{0.05}$As$_{2}$ powder, which is used as the flux in crystal growth. The powder XRD measurement for crushed crystals (polycrystal) was also carried out to check for possible phase decomposition. As shown in Fig. 2 (b), most of reflections could be indexed to the tetragonal structure with $a=b=$ 3.93 \AA\ at room temperature, except for several small peaks marked by stars from residual flux, which is very common in the self-flux growth method.

To determine the real chemical compositions of our single crystals, we selected several compounds with $x=$ 0.10, 0.175, 0.20, 0.225, 0.25 for inductively coupled plasma (ICP) analysis.
The segregation coefficient, namely the ratio between real concentration and nominal concentration, $K= C_s/C_l$ for Ni and Cr is about 0.84 and 0.71 in average, respectively, which is consistent with previous reports on BaFe$_{2-x}$Ni$_{x}$As$_{2}$ \cite{nni,rzhang,ychen}. In order to quantitatively compare with our earlier published results, we simply use the nominal composition to represent all samples in this paper.

\subsection{Antiferromagentism}

The antiferromagnetism was examined by neutron diffraction measurements. Figure 3 shows elastic ${\bf Q}$ scans along the $[H, H, 3]$ and $[0.5, 0.5, L]$ directions measured at the Rita-2 triple-axis spectrometer. In order to directly compare the magnetic intensity at each doping level and eliminate the $\lambda/2$ scattering, all data are normalized to $10^4$ monitor counts after subtracting the background scattering above $T_N$.
There is no magnetic intensity down to $T=$ 1.5 K for the $x=$ 0.20 compound measured at the TAIPAN triple-axis spectrometer, thus resulting zero net counts for ${\bf Q}$ scans [Fig. 3 (c) and (f)]. The instrument resolution for the Rita-2 shown as horizontal bars was estimated by using $\lambda/2$ scattering from the  $(1, 1, 6)$ nuclear Bragg peak above $T_N$ without filter, similar to the calculated values of the instrumental resolution ($R_H \sim 0.006$ r.l.u. for [H, H, 3] direction, $R_L \sim 0.024$ r.l.u. for [0.5, 0.5, L] direction) \cite{hqluo2012}. Although the real instrument resolution may be slightly larger than these values due to the supermirror guides in the incident beam of the Rita-2 spectrometer, the difference have negligible effect on samples with different doping levels measured at the same spectrometer. Experimentally, we find that all magnetic peaks shown in Fig. 3 are nearly resolution-limited and exhibit Gaussian distribution similar to the resolution function, suggesting long-ranged magnetic order in this system. The Gaussian fits to the peaks on zero backgrounds are shown by solid lines in Fig. 3 by the function: $I= I_0\exp[-(H-H_0)^2/(2\sigma^2)]$ for [H, H, 3] scans, or $I= I_0\exp[-(L-L_0)^2/(2\sigma^2)]$ for [0.5, 0.5, L] scans, where the full-width-half-maximum (FHWM) is $W=2\sqrt{2\ln{2}}\sigma$ in \AA$^{-1}$. After the correction from instrument resolution and Fourier transform of the Gaussian peaks from reciprocal space to real space \cite{hqluo2012,hjkang}, we can obtain the spin-spin correlation length $\xi_{ab}^{\mathrm{AF}}=8\ln(2)/\sqrt{W_H^2-R_H^2}/(2\pi\sqrt{2}/a)=2\sqrt{2}\ln(2)a/(\pi\sqrt{8\ln(2)\sigma^2-R_H^2})$ in \AA\ for in-plane scans along ${\bf Q}=[H, H, 3]$, and $\xi_{c}^{\mathrm{AF}}=8\ln(2)/\sqrt{W_L^2-R_L^2}/(2\pi/c)=4\ln(2)c/(\pi\sqrt{8\ln(2)\sigma^2-R_L^2})$ in \AA\ for out-of-plane scans along ${\bf Q}=[0.5, 0.5, L]$, where the lattice parameters $a \approx b \approx 3.94$ \AA, and $c = 12.90$ \AA\ at low temperature. Figure 6 (b) presents the doping dependence of spin correlation length in the $a$-$b$ plane  $\xi_{ab}$ and
along the $c$ axes $\xi_{c}$. Clearly, the spin-spin correlations at all doping levels are larger than 400 \AA, suggesting the long-ranged nature of the magnetic order. This is different from the BaFe$_{2-x}$Ni$_{x}$As$_{2}$, where the magnetic order becomes short-ranged when $x>0.08$ before being completely suppressed at $x=$ 0.108 [open circles in Fig. 6 (b)] \cite{hqluo2012,xylu2013}.

\begin{figure}[t]
\includegraphics[width=0.48\textwidth]{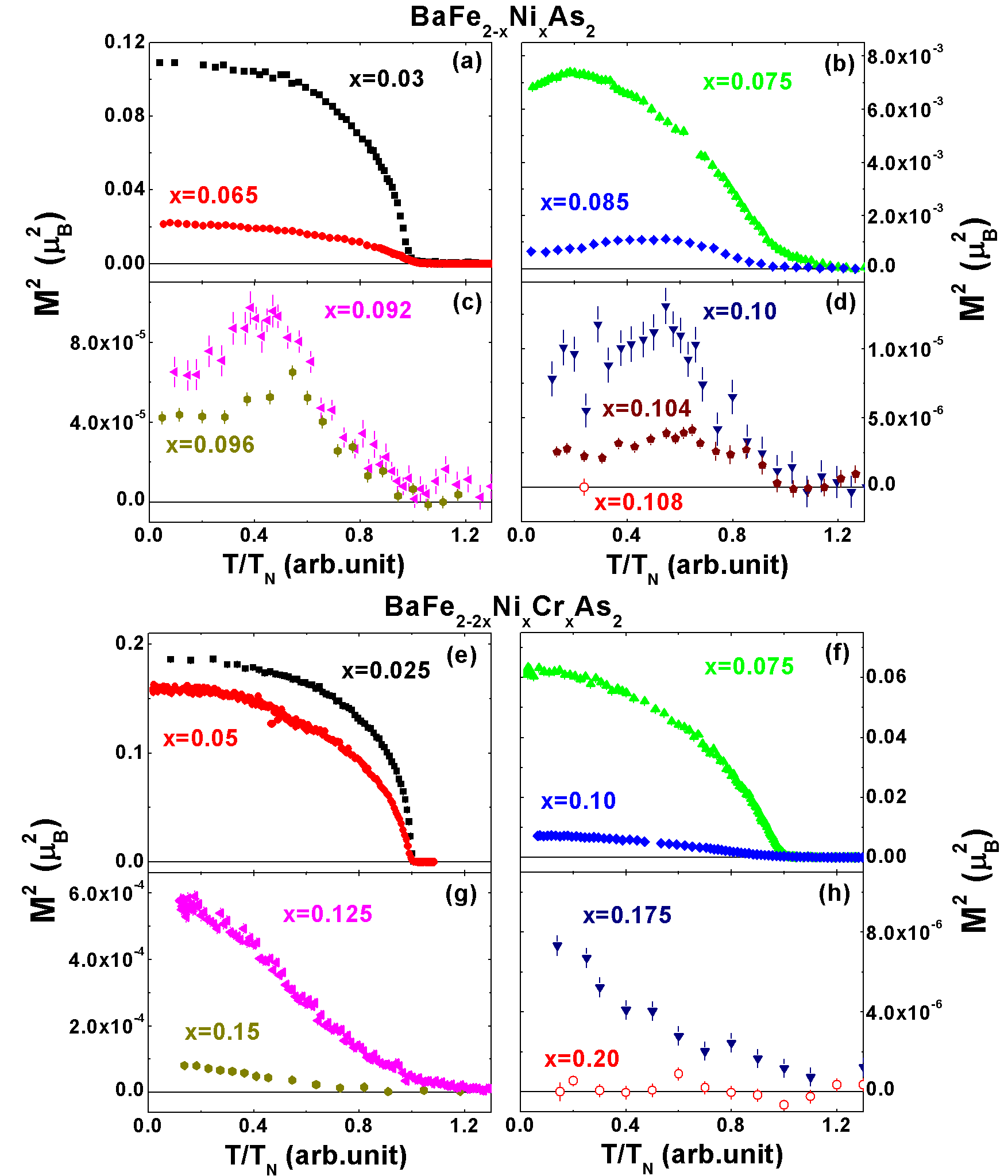}
\caption{ (Color online) Temperature dependence of the normalized
order parameter $M^2$ vs. $T/T_N$ for (a) -
(d)BaFe$_{2-x}$Ni$_{x}$As$_{2}$  and (e) - (h)
BaFe$_{2-2x}$Ni$_{x}$Cr$_{x}$As$_{2}$ . }
\end{figure}

\begin{figure}[t]
\includegraphics[width=0.4\textwidth]{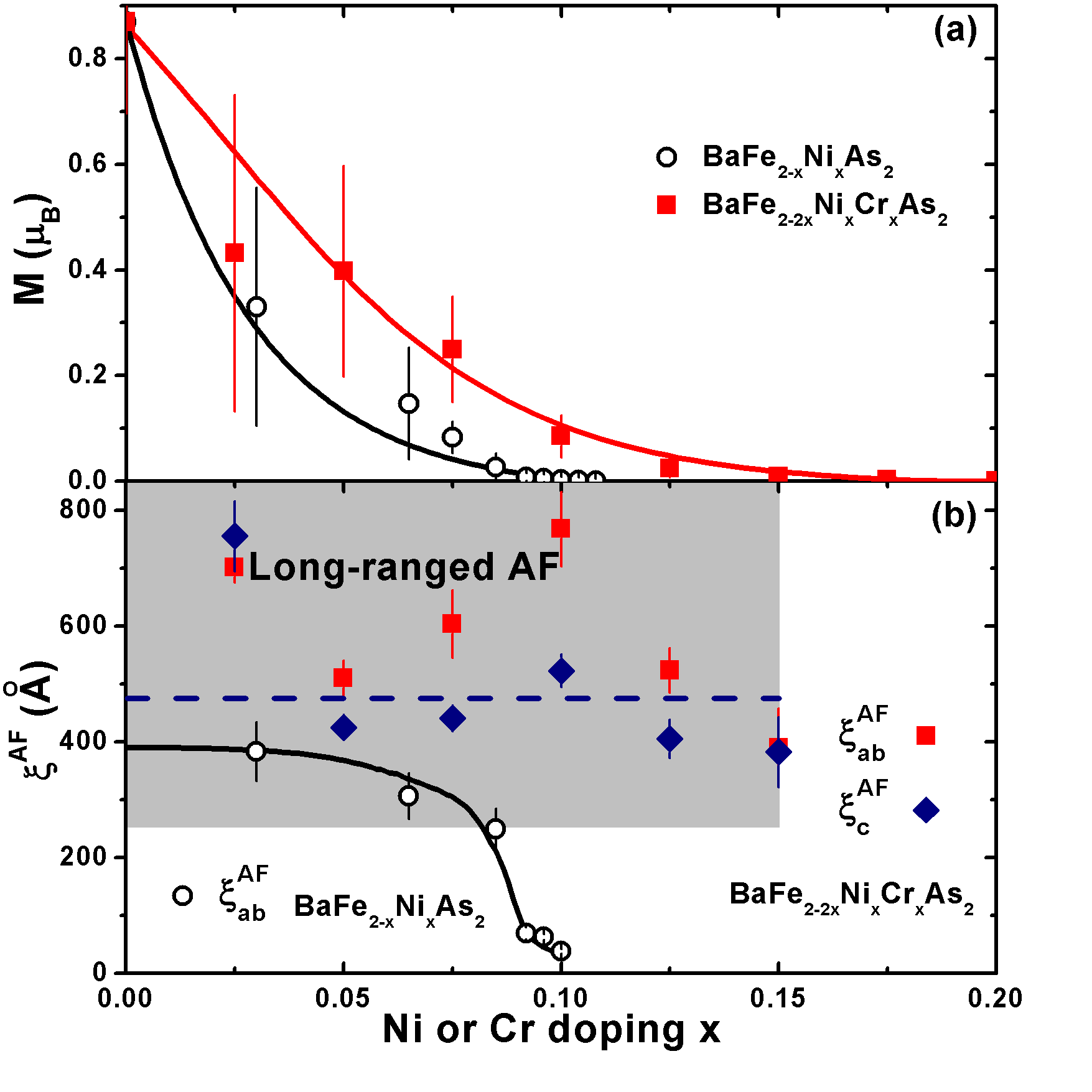}
\caption{ (Color online) Doping dependence of the (a) ordered moment
$M$ and (b) spin-spin correlation length $\xi_{AF}$ in
BaFe$_{2-2x}$Ni$_{x}$Cr$_{x}$As$_{2}$  and
BaFe$_{2-x}$Ni$_{x}$As$_{2}$ . }
\end{figure}

\begin{figure}[t]
\includegraphics[width=0.48\textwidth]{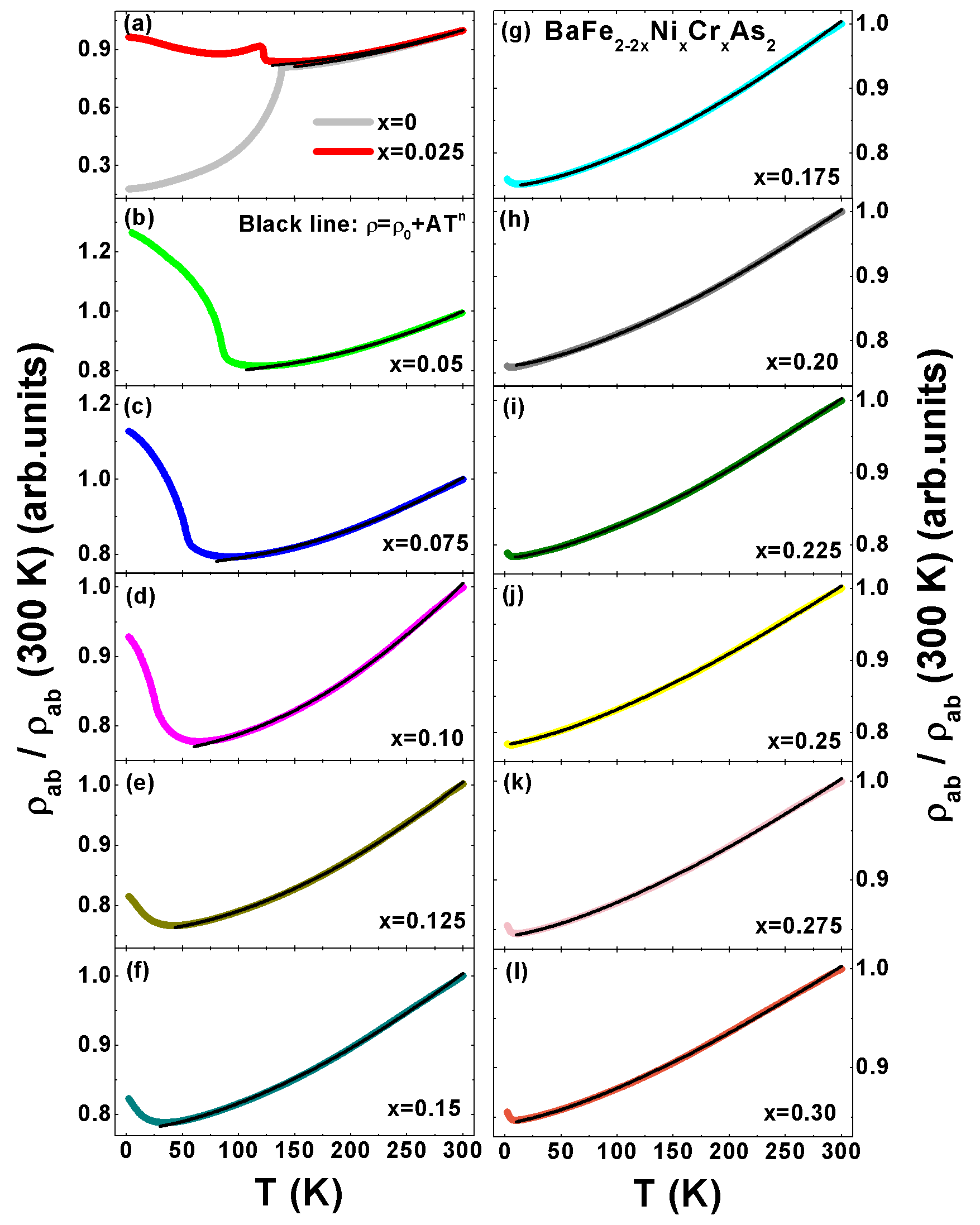}
\caption{ (Color online) Temperature dependence of in-plane
resistivity (normalized by the data at 300 K) for
BaFe$_{2-2x}$Ni$_{x}$Cr$_{x}$As$_{2}$. The black solid lines are
global fitting results by the model: $\rho=\rho_0+AT^n$ up to 300 K
in the normal state.}
\end{figure}

\begin{figure}[t]
\includegraphics[width=0.48\textwidth]{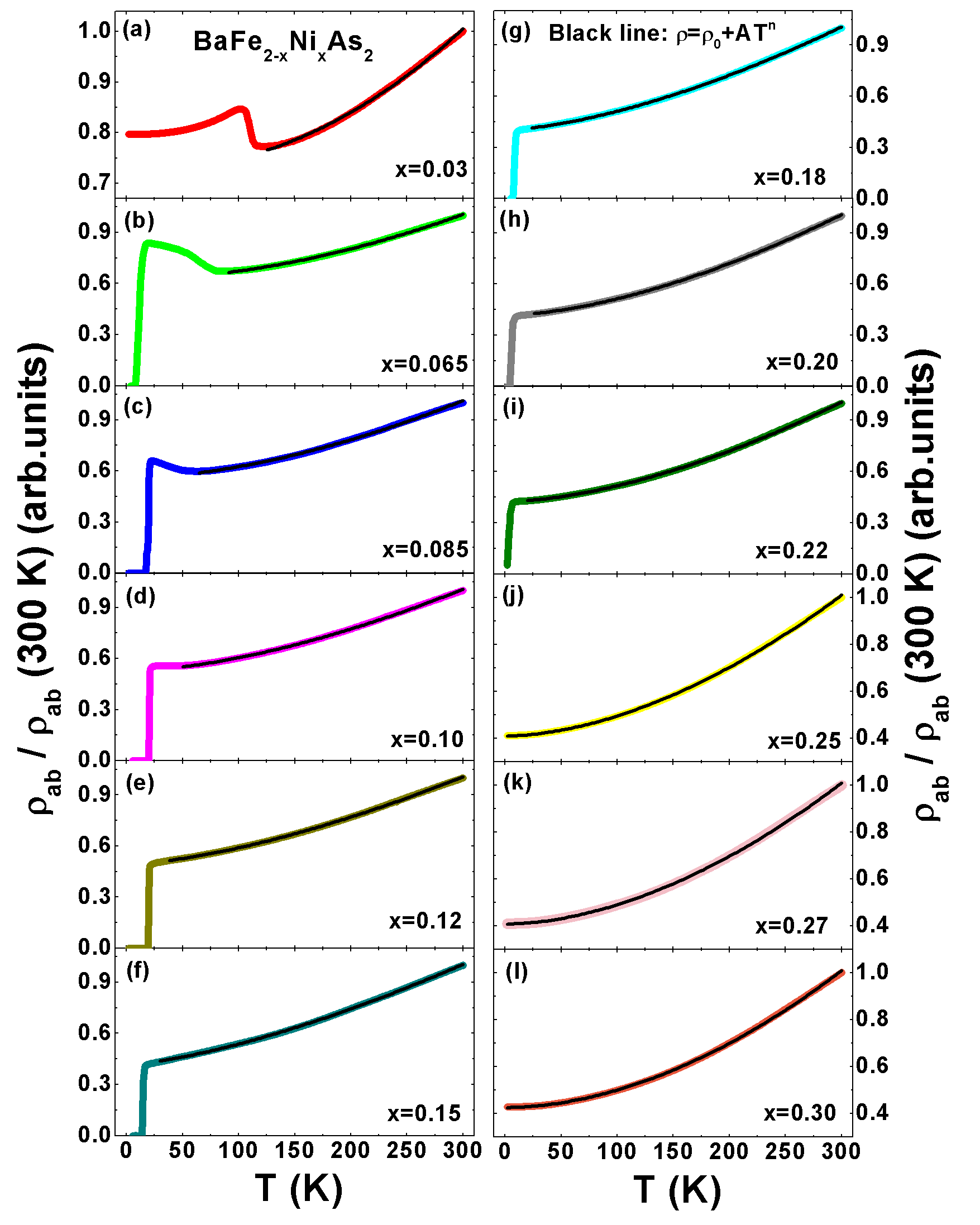}
\caption{ (Color online) Temperature dependence of in-plane
resistivity (normalized by the data at 300 K) for
BaFe$_{2-x}$Ni$_{x}$As$_{2}$. The black solid lines are global
fitting results by the model: $\rho=\rho_0+AT^n$ up to 300 K in the
normal state.}
\end{figure}

\begin{figure}[t]
\includegraphics[width=0.43\textwidth]{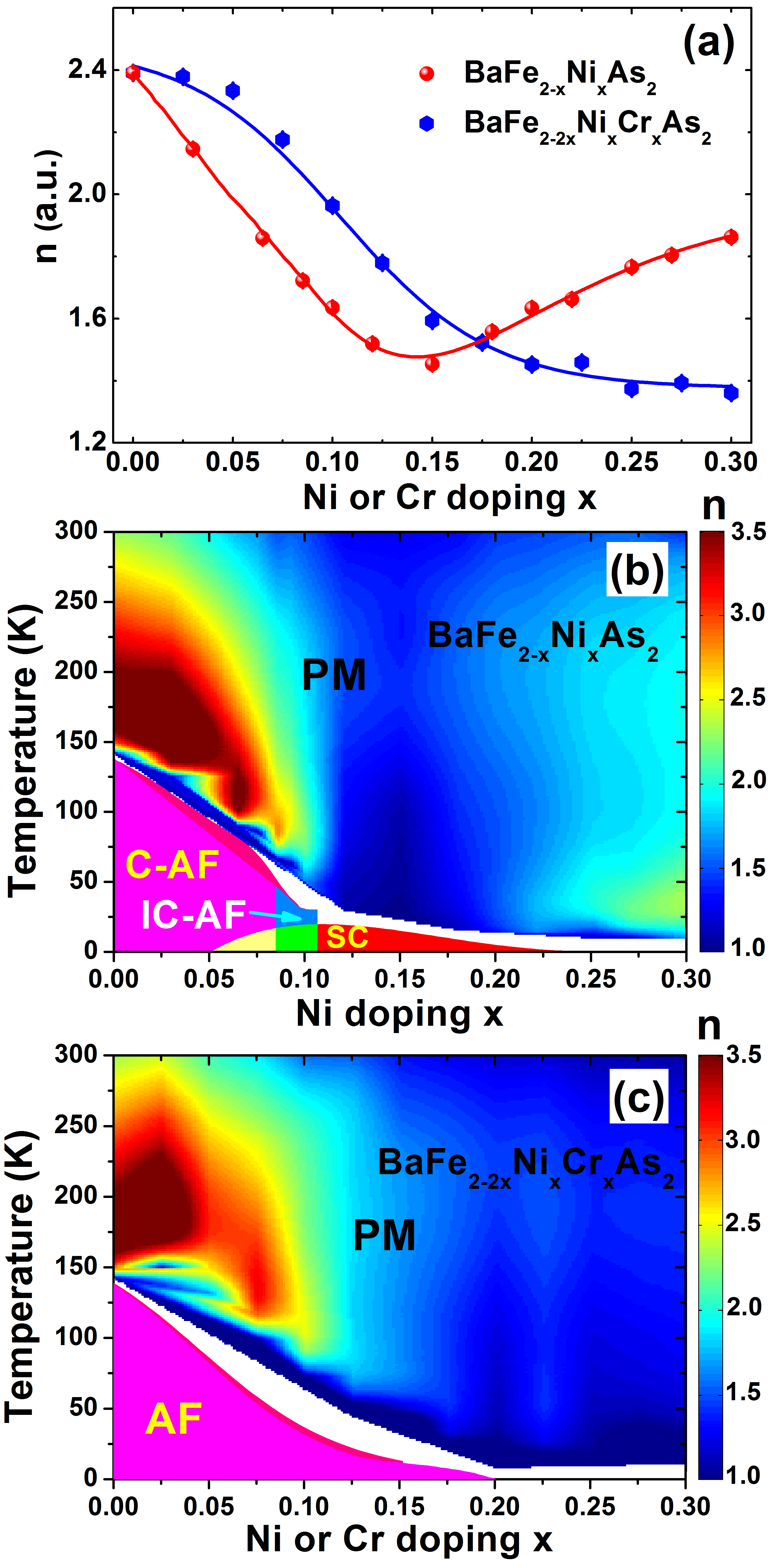}
\caption{(Color online) (a) Doping dependence of the exponent
 $n$ from global fitting results of Fig. 7 and Fig. 8. (b) and (c) The gradient color mapping for the
temperature and doping dependence of exponent $n$ for BaFe$_{2-x}$Ni$_{x}$As$_{2}$ and BaFe$_{2-2x}$Ni$_{x}$Cr$_{x}$As$_{2}$, respectively. }
\end{figure}

The temperature dependence of the intensity at $Q_{\mathrm{AF}}=(0.5, 0.5, 3)$ is shown in Fig. 4 (a) - (d), where the N\'{e}el temperature $T_N$ is marked by black arrows. For samples with large $x$ where an unambiguous determination of $T_N$ is difficult, we define $T_N$ as the cross point between the linear extrapolations of the decreasing part of the magnetic intensity upon warming
and the flat background at high-temperature. Here the data are also subtracted by the background above $T_N$ and normalized to $10^4$ monitor counts.  To determine the possible magnetic order for samples with high doping, we have co-aligned about 12 grams of the $x$= 0.20 and
23 grams of the $x= 0.25$ samples and measured them at the TAIPAN triple-axis spectrometer.
After counting about 1 hour for each temperature from 1.5 K to 15 K in 1 K per step at $ Q_{\mathrm{AF}}=(0.5, 0.5, 3)$, we find no evidence of static AF order for $x\ge 0.20$ [Fig. 4(d)]. The monotonically increasing intensity below $T_N$ for all magnetically ordered samples suggests that the AF ordered phase is not competing with other phases.

To examine the structure transition temperature $T_s$, we have measured the temperature dependence of nuclear Bragg reflection at the $Q=(1, 1, 0)$. As shown in Fig. 4 (e) - (h), an abrupt jump at $T_s$ arises from the neutron extinction release that occurs due to strain and domain formation related to the orthorhombic distortion \cite{xylu2}. For the $x=$ 0.10 sample, we see a clear kink at 31 K corresponding to structure transition.
For compounds with $x>0.125$, the extinction effect becomes unobservable due to much weaker orthorhombic distortion.

In elastic NS, the nuclear peak intensity $I_N$ and magnetic peak intensity $I_M$ are determined by \cite{gshirane}:
\begin{equation}
I_N=AN_N(2\pi)^2/V_N \times(\mid F_N(Q)\mid)^2/\sin(2\theta_N),
\end{equation}
and
\begin{equation}
I_M=AN_M(2\pi)^2/V_M \times(\mid F_M(Q)\mid)^2/(2\sin(2\theta_M)),
\end{equation}
where the number of atoms in magnetic unit cell is
$N_M=N_N/2=4$, the volumes of magnetic unit cell $V_M=2V_N$,
$F_N(Q)$ and $F_M(Q)$ are the structure factor and magnetic form
factor at wave vector $\mathbf{Q}$ with scattering angle $2\theta_N$
and $2\theta_M$, respectively. Here we have $F_M(Q)=pS\sum(-1)^ie^{iQd}$, where
$p=0.2659\times10^{-12}cm\times g\times f_M(Q)$ with $Fe^{2+}$ form
factor $f_M(Q)=Ae^{aQ^2/16\pi^2}+Be^{bQ^2/16\pi^2}+Ce^{cQ^2/16\pi^2}+D$,
$S$ is the magnetic moment along wave vector $\mathbf{Q}$,
$d$ is the spacing of wave vector $\mathbf{Q}$, $g$ factor is
assumed to be 2. The twinning effect from two kinds of magnetic domains is taken account in Eq. (2) by dividing the Lorentz factor $\mid F_M(Q)\mid^2/\sin(2\theta_M)$ by 2.
In principle, by comparing the integrated intensity between the nuclear and magnetic peaks in the form of Eq. (1) and (2),
we can estimate the static magnetic ordered moment via \cite{hqluo2012,slli2009}:
\begin{equation}
S=0.067\sqrt{I_M\sin{2\theta_M}/I_N\sin{2\theta_N}}\mid F_N\mid/\mid f_M\mid.
\end{equation}
Several factors have been carefully considered in such estimation:
(1).The neutron absorption may suppress the scattering intensity. In our experiments, we only use one piece of crystal for each measurement, the thickness is about 0.5 mm and almost the same for different dopings. Since our samples do not contain any elements with high neutron absorption, the effect of neutron absorption is negligible.
(2). To reduce neutron extinction effect, we use two weak nuclear peaks (1, 1, 0) (structural factor $F_N=1.31$) and (0, 0, 6) (structural factor $F_N=8.63$) above $T_s$ for normalizing magnetic peaks at $Q_{\mathrm{AF}}=(0.5, 0.5, 1)$ and $(0.5, 0.5, 3)$.
(3). To obtain reliable integrated magnetic/nuclear scattering in triple-axis NS experiments \cite{gshirane}, we have carried out $\theta-2\theta$ scans across the Bragg peaks. For samples with small mosaic (our sample mosaic is about 20$^\prime$)\cite{ychen}, this is a reliable way to obtain the integrated intensity. We can compare magnetic scattering from different $x$ as most of the experiments are carried out using Rita-2 with identical experimental setup.

Comparing the integrated intensity of magnetic peaks at $Q_{\mathrm{AF}}=(0.5, 0.5, 1)$ and $(0.5, 0.5, 3)$ and nuclear peaks at $Q=(1, 1, 0)$ and $(0, 0, 6)$, we estimate the ordered moments for samples with $x=0.025 - 0.175$ in Fig. 6 (a), where the values are taken as the statistical average from different combination of the scattering peaks, and the direction of the ordered moment is simply assumed to be the same as parent compound BaFe$_2$As$_2$, namely the longitudinal direction ${\bf Q}=[H, H, 0]$ for tetragonal lattice \cite{xylu2014,sdwilson,nqureshi}. The Cr doping enhances the magnetic ordered moment compared with those in BaFe$_{2-x}$Ni$_{x}$As$_{2}$,
where the moments are suppressed by superconductivity at low temperatures \cite{hqluo2012,rzhang}. For high dopings with $x=$ 0.20 and 0.25, the ordered moment $M$, if exist, is less than $10^{-3} \mu_\mathrm{B}$ at 1.5 K, which is out of the sensitivity of our measurements. The detailed temperature dependence of the ordered moment square $M^2$ (proportional to order parameters) for both systems is summarized in Fig. 5. The $M^2$ vs. $T$ in both systems behave similarly. The reduction of $M^2$ in BaFe$_{2-x}$Ni$_{x}$As$_{2}$ below $T_c$ is due to the effect of superconductivity \cite{hqluo2012,xylu2013}, which does not appear
in the non-superconducting BaFe$_{2-2x}$Ni$_{x}$Cr$_{x}$As$_{2}$ .

Based on the doping dependence of $T_s$ and $T_N$ in this system, we construct the phase diagram
of BaFe$_{2-2x}$Ni$_{x}$Cr$_{x}$As$_{2}$
in Fig. 1 (b).  For comparison, we also show the phase diagram of BaFe$_{2-x}$Ni$_{x}$As$_{2}$ in Fig. 1(a).
We find that BaFe$_{2-2x}$Ni$_{x}$Cr$_{x}$As$_{2}$ behaves similarly to Cr free samples for $x \leq 0.10$.
 However, the long-range AF order still persists for samples with 0.10 $\leq x \leq$ 0.175, and decreases with increasing $x$ until vanishing
around $x$= 0.20. These results are clearly different from the case in BaFe$_{2-x}$Ni$_{x}$As$_{2}$, where both $T_s$ and $T_N$ saturates around 30 K above optimal $T_c$= 20 K, and the magnetic order disappears in a first order fashion around $x$= 0.108.
In addition, the Cr substitution successfully eliminated the superconducting dome and short-ranged magnetic order in BaFe$_{2-x}$Ni$_{x}$As$_{2}$ system, which causes the avoided QCP around optimal superconducting doping $x$= 0.10.
These results suggest that Cr-doping is an effective way to study magnetism in iron pnictides without the complication of superconductivity.

\subsection{Resistivity}

We now discuss the in-plane resistivity measurements for all doping
levels of BaFe$_{2-2x}$Ni$_{x}$Cr$_{x}$As$_{2}$. Figure 7 show the normalized
resistivity $\rho_{ab}(T)/\rho_{ab}(300 \mathrm{K})$ from 2 K to 300 K for $x=$ 0 - 0.30. No
superconductivity is found down to 2 K for all $x$ \cite{rzhang}. The temperature dependence of $\rho_{ab}(T)$ is
metallic in the paramagnetic state, and the upturns at low
temperatures for lightly doped samples are attributed to the
structure and magnetic transition, while the small kink for $x\geq$
0.2 may be from impurity scattering. To fully understand the
behavior of $\rho_{ab}(T)$ in the paramagnetic state, we first
perform a global fit for the data from $T_s+20$ K to 300 K by
the empirical model \cite{hqluo2012,arosch,yluo,rzhou,Analytis}:
\begin{equation}
\rho=\rho_0+AT^n,
\end{equation}
where $\rho_0$ is a constant, $A$ is amplitude and $n$ is the exponent.
Typically, a $n=$ 2 power law is expected of a Fermi liquid, while a
 $n<$ 2 power law means non-Fermi-liquid behavior, and a QCP usually corresponds to $n=$ 1
as a function of tuning parameter such as doping, pressure or magnetic field \cite{Analytis,yluo,racooper,jganalytis}.
In most cases, such fits are performed at low temperature for the domination of electron-electron interactions.
However, due to the limited effects on the Debye temperature from Ni or Cr doings less than 15\% in the system \cite{bzeng}, it is reasonable to roughly compare the exponent $n$ up to 300 K for different doping with similar contributions from phonon.

The fitting results are shown in solid black lines in Fig. 7.
Similar analysis is also done for BaFe$_{2-x}$Ni$_{x}$As$_{2}$
system, as separately shown in Fig. 8 and agree with the original data very
well. The exponent $n$ from global fits for different $x$ are summarized in
Fig. 9 (a). The non-Fermi-liquid behavior with $n<$ 2 in
BaFe$_{2-x}$Ni$_{x}$As$_{2}$, occurs above $x=$ 0.05, concurrent
with the appearance of doping induced superconductivity. The value
of $n$ at $x=$ 0.15 reduces to about 1.5. Upon doping in
BaFe$_{2-2x}$Ni$_{x}$Cr$_{x}$As$_{2}$, $n$ continuously decreases
with increasing $x$ and saturates to around 1.4 for $x\geq$ 0.20.
Therefore, although the Cr impurities have quite limited effects on
$T_s$ and $T_N$ except for suppressing superconductivity, the normal
state transport properties are different from pure Ni-doped
material. We note that the minimal $n$ in
BaFe$_{2-x}$Ni$_{x}$As$_{2}$ is 1.5 in the overdoped regime,
different from the expectation of a standard QCP point. This is
consistent with neutron diffraction and XRD results of no QCP near
optimal superconductivity \cite{hqluo2012,xylu2013}. Similarly,
there is also no dramatic anomaly in $n$ at any doping level
of BaFe$_{2-2x}$Ni$_{x}$Cr$_{x}$As$_{2}$, consistent with the
absence of a QCP near $x=0.20$, even though the $T_N$ of this compound
is indeed below 2 K.

To exclude the phonon contributions at high temperatures, an alternative way of analyzing resistivity data is to
simply deduce the temperature dependence of $n$ from the slope of the curve $\ln(\rho_{ab}-\rho_0)$ as a function of $\ln T$
using Eq. 4.  Here the $\rho_0$ is determined by local fits from $T_s+20$ K to $T_s+70$ K. Such process requires high quality data of $\rho_{ab} (T)$.
Figure 9 (b) and (c) show the temperature and doping dependence of $n$ in the whole phase diagram.
The outcome further confirms the global fitting results in Fig. 9 (a), where the $n<2$ behavior mainly exists in BaFe$_{2-2x}$Ni$_{x}$Cr$_{x}$As$_{2}$ over $x=$ 0.10 and in the middle zone with 0.10 $\leq x \leq$ 0.20 of BaFe$_{2-x}$Ni$_{x}$As$_{2}$, respectively. We note that the features in Fig. 9 (b) are similar to BaFe$_{2}$(As$_{1-x}$P$_x$)$_2$ system where the presence of a QCP has been suggested \cite{Shibauchi},
although BaFe$_{2-x}$Ni$_{x}$As$_{2}$ is in the paramagnetic state above $x=0.108$ \cite{xylu2013}.

\begin{figure}[t]
\includegraphics[width=0.43\textwidth]{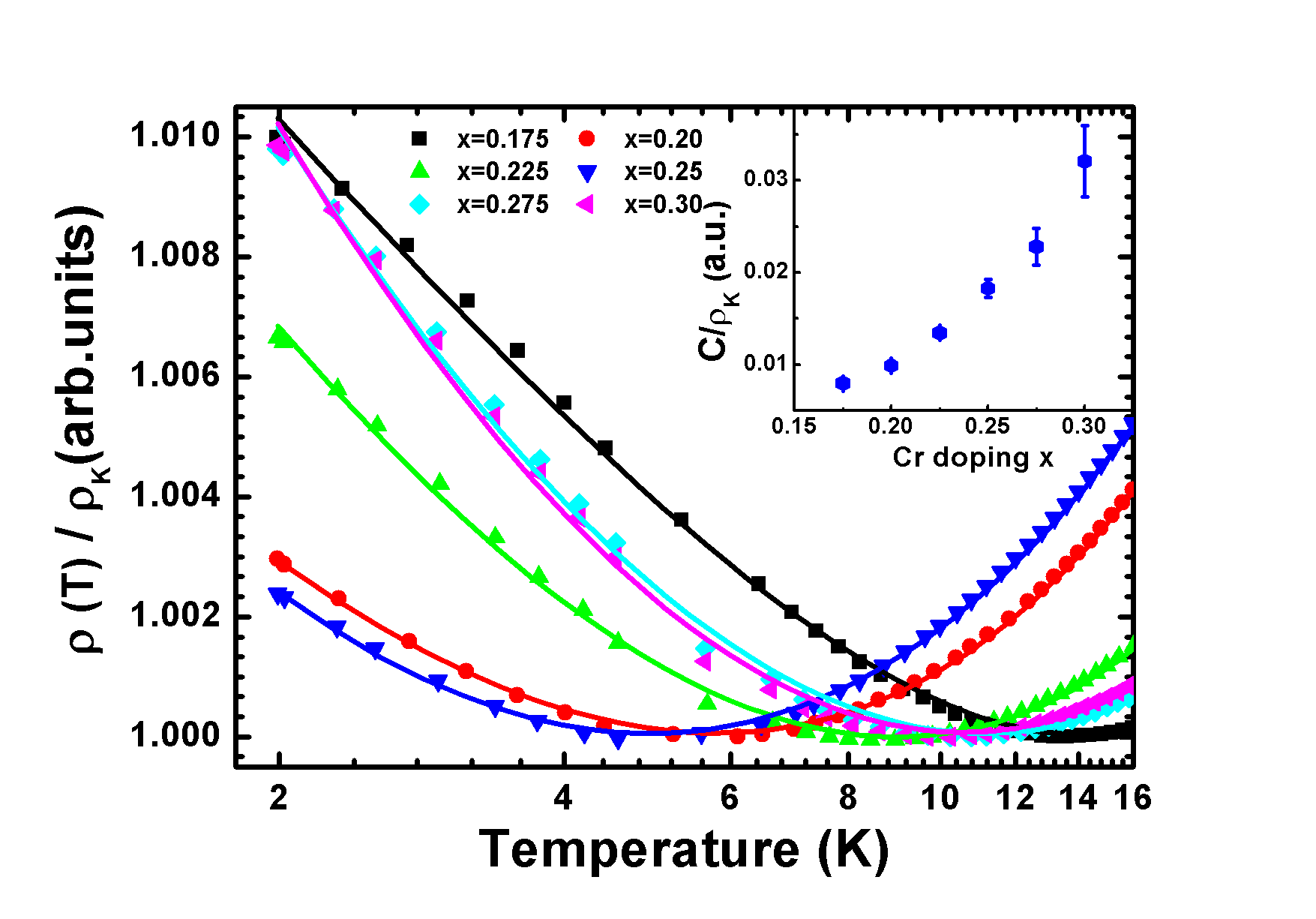}
\caption{ (Color online) Logarithmic temperature dependence of
resistivity for the compounds with $x=$ 0.175 - 0.30. For clarity,
all data are normalized by the resistivity at the kink temperature
$\rho_K$. The solid
 lines are fitting results by Eq. 5.}
\end{figure}

\begin{figure}[t]
\includegraphics[width=0.42\textwidth]{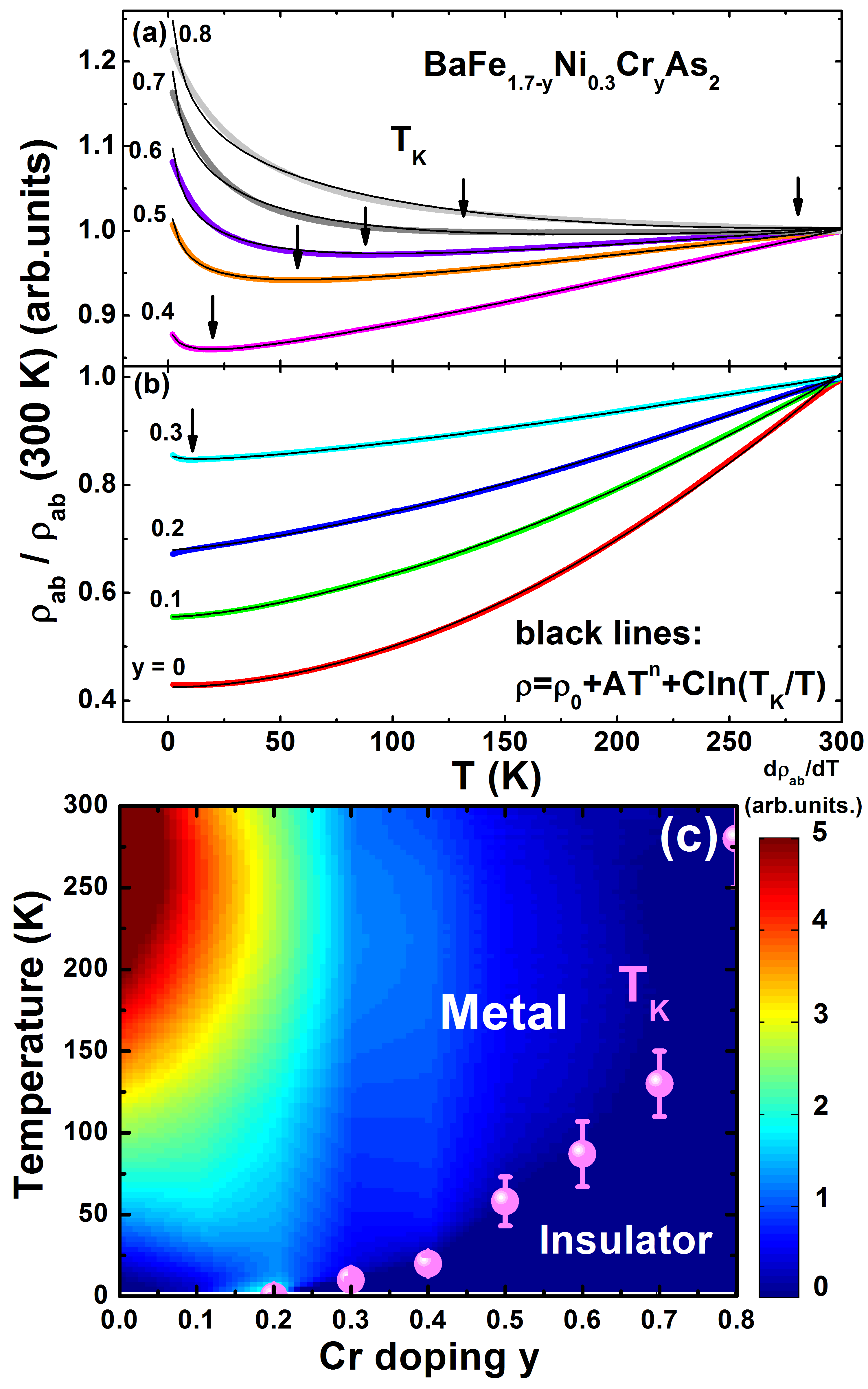}
\caption{ (Color online) (a) and (b) Temperature dependence of
in-plane resistivity for BaFe$_{1.7-y}$Ni$_{0.3}$Cr$_{y}$As$_{2}$
with different Cr doping $y$.  The black solid
 lines are fitting results by Eq. 5. (c) The gradient color mapping for the
temperature and doping dependence of the first order differential of
resistivity $d\rho_{ab}/dT$ of
BaFe$_{1.7-y}$Ni$_{0.3}$Cr$_{y}$As$_{2}$, where the sign change
temperature (kink temperature) $T_K$ marks the zone boundary between
metallic and insulating state.}
\end{figure}

\begin{figure}[t]
\includegraphics[width=0.35\textwidth]{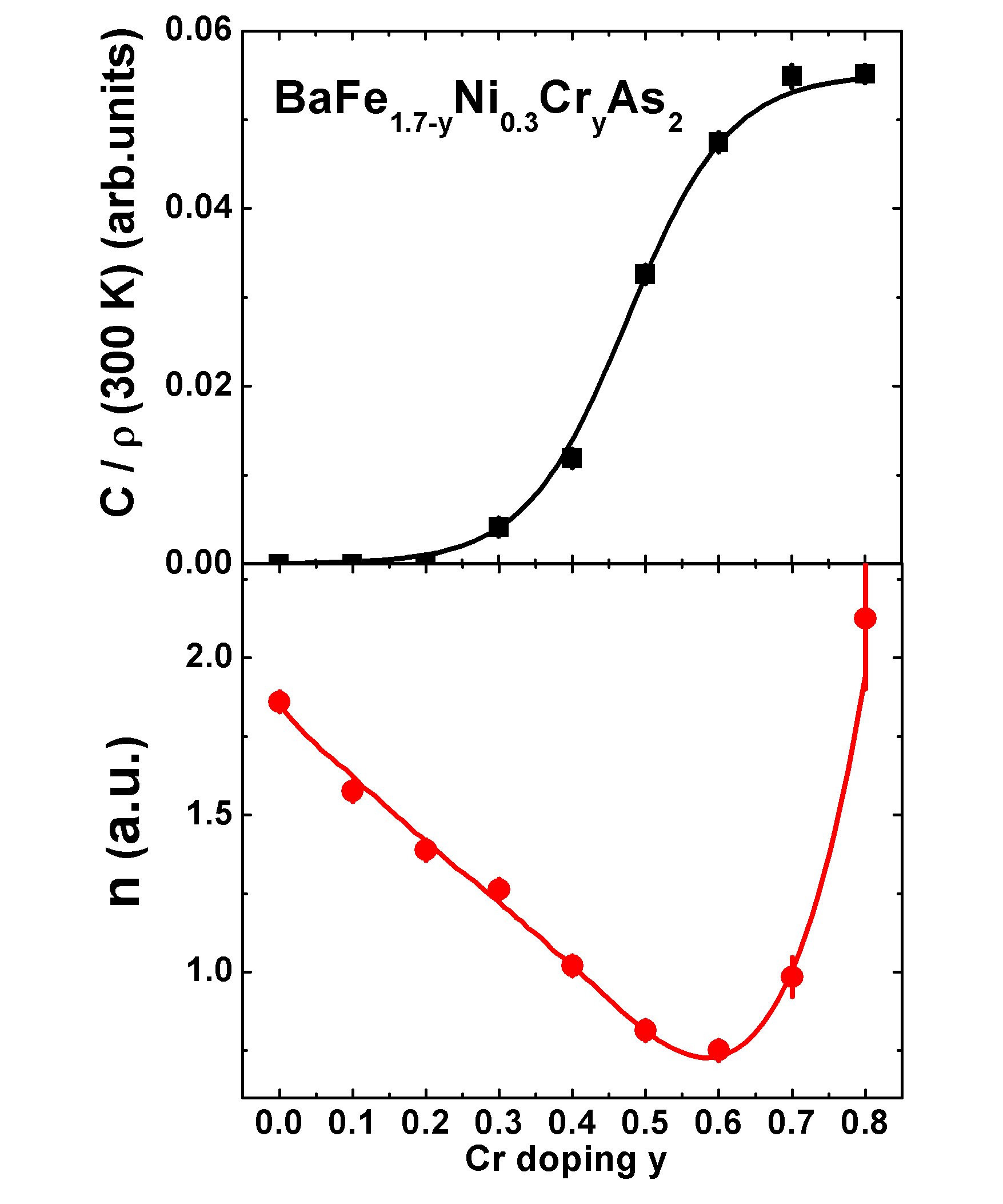}
\caption{ (Color online) Doping dependence of fitting parameters
$C/\rho (300 K)$ and $n$ from Fig. 11.}
\end{figure}

\subsection{Impurity effect}
By successfully eliminating the superconducting dome in BaFe$_{2-x}$Ni$_{x}$As$_{2}$ via doping equivalent Cr and Ni, we obtain a simple phase diagram in BaFe$_{2-2x}$Ni$_{x}$Cr$_{x}$As$_{2}$ system, where the
$T_N$ of the material is systematically suppressed to below 2 K around $x=0.20$.
Usually, such a phase diagram would be a strong indication for a magnetic QCP,
since the spin correlation length of the AF ordered phase is long-ranged for all doping levels.
However, the transport properties of the system reveal no anomaly typically
associated with a QCP.  Future inelastic NS experiments for samples near $x=0.20$ should be able to determine if
energy and temperature are scaled, as is the case for a magnetic QCP \cite{schroeder}.
 From the results in Fig. 1 and Fig. 9, one could argue that the non-Fermi-liquid behavior may not be intimately associated with the disappearance of magnetic order in these systems.
On the other hand, for a multiband material such as iron pnictides, it is not clear if transport measurements alone can indeed prove the presence of a QCP.
 By introducing Cr impurity into BaFe$_{2-x}$Ni$_{x}$As$_{2}$ to remove superconductivity, the original magnetic phase boundary
has been changed possibly due to the impurity effect of Cr doping. Even in the Cr free samples, the Ni doping can cause anisotropic transport properties \cite{Bohmer,Allan,Ishida}.

To better understand the effect of impurity scattering, we note the
presence of a small upturn in $\rho_{ab}(T)$ below 10 K even in the
highly doped compounds without AF order. Because the Cr ion is
magnetic, by doping Cr impurity into a metallic system, additional
Kondo scattering may be induced. If this is indeed the case, one may expect
a logarithmic dependence of $\rho(T)$ at low temperature, since the
temperature dependence of the resistivity including the Kondo effect
can be written as: $\rho(T)=\rho_0+AT^2+C\ln(T_K/T)+BT^5$, where
$A$,$B$ and $C$ are constant, and $T_K$ is Kondo temperature
\cite{jkondo}.  Moreover, a weak localization effect induced by
local impurity in the two dimensional electron system may also give
rise to a $\ln T$ dependence of resistivity at low temperature
\cite{palee1985}, which is common in the chemical substituted
copper-oxide high temperature superconductors \cite{hqluo2014}.
Interestingly, by plotting the normalized resistivity below 10 K in
the logarithmic axis of temperature in Fig. 10, we have found a well
defined linear region of $\rho(T)$ as a function of $\ln T$ for
$x\geq$ 0.175. We then fit the data by the formula:
\begin{equation}
\rho(T)=\rho_0+AT^n+C\ln(T_K/T),
\end{equation}
where $T_K$ is fixed as the kink temperature of $\rho_{ab}(T)$,
namely, the sign change in temperature of $d\rho_{ab}/dT$. The fitting
results as shown by solid lines in Fig. 10 agree very well with the
data, and an increase of normalized $C/\rho_K$ upon doping $x$
is found (insert of Fig. 10). Thus the impurity scattering may be
proportional to the Cr doping level $x$. To further confirm this
idea, we then fix the Ni concentration as $x=$ 0.30 and change Cr
from $y=$ 0 to $y=$ 0.8 in BaFe$_{1.7-y}$Ni$_{0.3}$Cr$_{y}$As$_{2}$
system. The normalized in-plane resistivity is show in Fig. 11 (a) and (b). For small Cr impurity with
$y\leq$ 0.2, the samples are metallic. Starting from $y=$ 0.3, a
clear upturn in $\rho_{ab}(T)$ emerges and diverges quickly at low
temperatures, finally dominates in the whole temperature range up to
300 K for $y=$ 0.8, suggesting a metal-to-insulator crossover in
high Cr impurity doped compounds. The color mapping of
$d\rho_{ab}/dT$ in Fig. 11 (c) confirms this conclusion, where the blue zone marks the negative $d\rho_{ab}/dT$.

We also fit the resistivity data using Eq. 5 and obtain the parameters shown in
Fig. 12. It seems that $C$ steeply grows up after $x=$ 0.3 for strong impurity effects. Surprisingly, the exponent $n$
almost linearly decrease from about 2 to less than 1 and finally
increase back to around 2. Such abnormal behaviors are similar to
Cr-free BaFe$_{2-x}$Ni$_{x}$As$_{2}$ system in a different way of ionic
substitutions. Thus it strongly suggests the impurity scattering
cannot be ignored in the transport properties in these materials,
which may also affect the interpretation of the resistivity data.
Therefore, the non-Fermi-liquid feature associated with $n<$ 2 may be
not a direct evidence for the presence of a magnetic QCP.

\section{Conclusions}
In conclusion, we use elastic neutron scattering and electric transport to study the antiferromagnetic and paramagnetic state behaviors of non-superconducting iron pnictide
BaFe$_{2-2x}$Ni$_{x}$Cr$_{x}$As$_{2}$. We find that the long-range AF order terminates at $x=$ 0.20 with a systematic decrease of ordered moment and N\'{e}el temperature with increasing $x$.
Although the resistivity results suggest non-Fermi-liquid behavior above $x=$ 0.10, there is still no direct evidence for a magnetic QCP, while the transport properties are significant affected by Kondo
scattering or weak localization effect from Cr impurity.

\section{Acknowledgments}

The work at IOP, CAS is supported by the National Natural Science Foundation
of China (Projects No. 11374011 and No. 91221303), the Ministry of Science and Technology of China
(973 projects: 2011CBA00110 and 2012CB821400), and The Strategic Priority Research Program
(B) of the Chinese Academy of Sciences (Grant No.
XDB07020300). R. Z., X. L. and H. L. acknowledge Project No. 2013DB03 supported by NPL, CAEP.
The work at Rice is supported by NSF DMR-1362219, DMR-143606, and the Robert A.
Welch Foundation Grant No. C-1839. Part of this work is based on experiments performed at the Swiss spallation neutron source SINQ, Paul Scherrer Institute, Villigen, Switzerland. The authors thank the
fruitful discussion with Tao Xiang, and the great
help on the crystal growth and characterization from Zhaoyu Liu, Wenliang Zhang, Tao Xie, Lihong Yang, Jun Zhu, Cong Ren and Xingjiang Zhou.


\end{document}